\documentclass[twocolumn,american]{iopart}

\makeatletter
\usepackage{iopams}
\usepackage{setstack}

\makeatother

\usepackage{babel}

\begin{document}

\rapid{Pure 4-geometry of quantum magnetic spin matter from Kondo effect }

\author{Torsten Asselmeyer-Maluga}

\address{German Aerospace center, Rutherfordstr. 2, 12489 Berlin, Germany }

\ead{torsten.asselmeyer-maluga@dlr.de}

\author{Jerzy Kr\'ol}

\address{University of Silesia, Institute of Physics, ul. Uniwesytecka 4,
40-007 Katowice, Poland}

\ead{iriking@wp.pl}

\begin{abstract}
We determine a smooth Euclidean 4-geometry on $\mathbb{R}^{4}$ from
quantum interacting spin matter like in the multichannel Kondo effect.
The CFT description of both: the $k$-channel Kondo effect of spin
magnetic impurities quantum interacting with spins of conducting electrons
and exotic smooth $\mathbb{R}_{k}^{4}$, by the level $k$ WZW model
on $SU(2)$, indicates the relation between smooth $\mathbb{R}^{4}$'s
and the quantum matter. We propose a model which shows: exotic smooth
$\mathbb{R}_{k}^{4}$ generates fermionic fields via the topological
structure of Casson handles and when this handle is attached to some
subspace $A$ of $\mathbb{R}^{4}$ these fermions represent electrons
bounded by the magnetic impurity. Thus the Kondo bound state of $k$
conducting electrons with magnetic impurity of spin $s$ is created
like in the low temperature Kondo effect. Then the quantum character
of the interactions is encoded in 4-exoticness. The complexity as
well the number of Casson handles correspond to the number of channels
in the Kondo effect. When the smoothness structure is the standard
one, no quantum interactions are carried on by standard $\mathbb{R}^{4}$. 
\end{abstract}
\submitto{\JPA}
\maketitle
Einsteins equation describes the balance of the density of gravity
with density of energy and matter in 4-dimensional spacetime. The
density of gravity is the curvature of smooth 4-manifold. All kinds
of matter and energy have their contributions to $T_{\mu\nu}$ hence
affect the metric structure of spacetime. Quantum matter contributes
by the averaged energy-momentum tensor. This shows that there is no
substantial difference between quantum and classically described matter
provided the densities entering $T_{\mu\nu}$ are the same. The metric
field remains classical also in the quantum regime of matter. Such
a classical universality would be an attractive feature if general
relativity (GR) is consistent with quantization of interactions prescribed
by the Standard Model (SM) of particle physics. However, this is not
the case: quantum field theory (QFT) faces problems even with its
formulation on dynamical curved backgrounds. On the other hand quantization
of GR is not consistent in dimension 4 with SM. Thus leaving the classical
Einsteins equation at the fundamental level is not a solution and
the attempts to quantize these theory fail in dimension 4. Since,
classical GR equivalently deals with geometry of 4-smooth manifolds,
the above problem is the fate of a smooth 4-geometry in a description
of a quantum regime of gravity or the relation of smooth 4-geometry
with quantum matter in general spacetimes. Two major possibilities
follow: a quantum spacetime analogue is not a \emph{smooth} 4-manifold
any longer and, because of that, it can be related with quantum matter,
or spacetime is still represented by a smooth 4-manifold though this
time it has to be related to quantum matter in a different way than
by averaged energy densities. 

This letter explains how quantum interacting spin matter, as in the
multichannel Kondo effect, relates to smooth Euclidean 4-geometry.
We propose a model where the quantum interacting spin matter is indeed
derivable from the smooth Euclidean 4-manifold structure alone. Thus
the classical, smooth geometry of 4-manifolds can naturally overlap
with the quantum regime of interacting matter fields. Thus the dichotomy
between the classical and the quantum regime (or between the smooth
4-geometry of manifolds and non smooth models) is overpassed because
smooth 4-geometry is essentially \emph{the same} as quantum interacting
spin matter. The physical Kondo effect holds in a variety of alloys
at low temperatures hence certainly not in the regime of Planck energies
or quantum gravity. This kind of relation of smooth 4-manifolds with
quantum matter appears as new phenomenon which has never been discussed
before. In the following we will understand a 4-geometry as smooth
exotic $\mathbb{R}^{4}$. We refer the reader to the excellent textbooks,
physical as well mathematical, relating exotic smooth $\mathbb{R}^{4}$
\cite{Asselmeyer2007,GomSti:1999}. Every exotic $\mathbb{R}^{4}$
is a smooth 4-manifold which is topologically the same as (homeomorphic
to) the standard smoothness structure as the topological product of
the four coordinate axes. However, every exotic $\mathbb{R}^{4}$
is smoothly distinct from (or nondiffeomorphic to) the standard $\mathbb{R}^{4}$.
Therefore we call these spaces \emph{nonstandard} or \emph{exotic
$\mathbb{R}^{4}$}. Every Riemannian 4-manifold is locally described
by the standard $\mathbb{R}^{4}$. It follows that every exotic $\mathbb{R}^{4}$
can not be covered by a single coordinate patch, which is a standard
$\mathbb{R}^{4}$. Thus a nontrivial metric exists on an exotic $\mathbb{R}^{4}$
(or one cannot define a constant smooth metric). Hence a form of gravity
might be present on this $\mathbb{R}^{4}$ \cite{BraRan:93,Bra:94b,Sladkowski2001}.
The existence of nonstandard smoothness of $\mathbb{R}^{n}$ is assigned
exclusively to dimension $n=4$. In fact there is a continuum of inifinitely
many different exotic $\mathbb{R}^{4}$ but only one smooth standard
$\mathbb{R}^{n}$ for any other dimension $n\neq4$. 

GR shows that matter/energy is linked with 4-geometry of smooth manifolds.
In our model quantum interacting spin-magnetic matter is also linked
with 4-geometry of the smooth, though nonstandard, $\mathbb{R}^{4}$.
We will apply our model to the case of the multichannel Kondo effect
with 2-components spins. Exotic $\mathbb{R}^{4}$'s are Euclidean
4-manifolds. The existence of Minkowski metric on exotic $\mathbb{R}^{4}$
is a strong limitation \cite{Sladkowski2001}. This would suggest
that the above mentioned link could have a rather nonperturbative
description at the dual limit of some quantum field theory (cf. \cite{Witten85}).
This important issue, however, will be addressed separately.

{\bf CFT, magnetic spin-$s$ impurities and exotic $\mathbb{R}^{4}$}.
Kondo proposed a simple phenomenological Hamiltonian \cite{Affleck95}: 

\begin{equation}
H=\sum_{\overset{\rightarrow}{k},\alpha}\psi_{\overset{\rightarrow}{k}}^{+\alpha}\psi_{\overset{\rightarrow}{k}\alpha}\epsilon(k)+\lambda\overset{\rightarrow}{S}\cdot\sum_{\overset{\rightarrow}{k},\overset{\rightarrow}{k'}}\psi_{\overset{\rightarrow}{k}}^{+}\frac{\overset{\rightarrow}{\sigma}}{2}\psi_{\overset{\rightarrow}{k'}}\,.\label{eq:Hamiltonian-Kondo}\end{equation}
explaining the growth of the resistivity $\rho(T)$ in some metals
when the temperature $T$ is lowering below the Kondo temperature
$T_{K}$ in the presence of magnetic spin-$s$ impurity. Here $\psi$
is the anihilation operator for the conduction electron of spin $\alpha$
and momentum $\overset{\rightarrow}{k}$, the antiferromagnetic interaction
term is that of spin impurity $\overset{\rightarrow}{S}$ with spins
of conducting electrons, at $\overset{\rightarrow}{x}=0$. From this
Hamiltonian one can derive in the Born approximation $\rho(T)\sim\left[\lambda+\nu\lambda^{2}\ln\frac{D}{T}+...\right]^{2}$where
the second term is divergent in $T=0$. The Hamiltonian (\ref{eq:Hamiltonian-Kondo})
is derivable from the more microscopic Anderson model \cite{Anderson1961}.
The Kondo antiferromagnetic coupling appears as the tunneling of electrons
screening the spin impurity, hence it is a pure quantum process (see
eg. \cite{Potok2007}). 

The exact low $T$ behavior was insightfully and exactly reformulated
by Affleck and Ludwig by the use of boundary conformal field theory
(BCFT) techniques \cite{AffleckLudwig91,AffleckLudwig94,Affleck95,AffleckLudwig93,Potok2007}
which is also the base for working out the connection with smooth
4-geometry.

Namely, given an integer class $h=k[\:]\in H^{3}(S^{3},\mathbb{Z})$
where $[\:]$ denotes the generator of $H^{3}(S^{3},\mathbb{Z})$.
To this class one can associate the corresponding WZW model on $S^{3}=SU(2)$
at integer level $k$. Changing the class $k[\:]\to l[\:]$ results
in the change of the level $k$ of the WZW model. The Kac-Moody algebra
$SU(2)_{k}$ is spanned on 3-components currents $\overset{\rightarrow}{{\cal J}}_{n}$,
$n=...-2,-1,0,1,2,...$:

\begin{equation}
[{\cal J}_{n}^{a},{\cal J}_{m}^{b}]_{k}=i\epsilon^{abc}{\cal J}_{n+m}^{c}+\frac{1}{2}kn\delta^{ab}\delta_{n,-m}\:.\label{eq:KMk}\end{equation}
Next we decompose the currents $\overset{\rightarrow}{{\cal J}}_{n}$
as $\overset{\rightarrow}{{\cal J}}_{n}=\overset{\rightarrow}{J}_{n}+\overset{\rightarrow}{S}$
such that $\overset{\rightarrow}{J}_{n}$ obey the same Kac-Moody
algebra, i.e. $[J_{n}^{a},J_{m}^{b}]_{k}=i\epsilon^{abc}J_{n+m}^{c}+\frac{1}{2}kn\delta^{ab}\delta_{n,-m}$
and usual relations for $\overset{\rightarrow}{S}$, i.e. $[S^{a},S^{b}]=i\epsilon^{abc}S^{c}$,
$[S^{a},J_{n}^{b}]=0$. From the point of view of field theories describing
the interacting currents with spins, $\overset{\rightarrow}{{\cal J}}_{n}$
corresponds to the effective infrared fixed point of the theory of
interacting spins $\overset{\rightarrow}{S}$ with $\overset{\rightarrow}{J}_{n}$where
the coupling constant $\lambda$ is taken as $\frac{2}{3}$ for $k=1$.
The interacting Hamiltonian of the theory for $k=1$ reads: 

\begin{equation}
H_{s}=c\left(\frac{1}{3}\sum_{-\infty}^{+\infty}\overset{\rightarrow}{J}_{-n}\cdot\overset{\rightarrow}{J}_{n}+\lambda\sum_{-\infty}^{+\infty}\overset{\rightarrow}{J}_{n}\cdot\overset{\rightarrow}{S}\right)\:.\label{eq:Action-spin-H}\end{equation}
For $\lambda=\frac{2}{3}$ one completes the square and the algebra
(\ref{eq:KMk}) for the currents $\overset{\rightarrow}{{\cal J}}_{n}$
follows., Then the new Hamiltonian where $\overset{\rightarrow}{S}$
is now effectively absent (still for $k=1$) is given by $H=c'\sum_{-\infty}^{+\infty}\left(\overset{\rightarrow}{{\cal J}}_{-n}\cdot\overset{\rightarrow}{{\cal J}}_{n}-\frac{3}{4}\right)\:.$

A similar procedure holds for arbitrary integer $k$ where the spin
part of the Hamiltonian reads: $H_{s,k}=\frac{1}{2\pi(k+2)}\overset{\rightarrow}{J}^{2}+\lambda\overset{\rightarrow}{J}\cdot$$\overset{\rightarrow}{S}\delta(x)$
and the infrared effective fixed point is now reached for $k=\frac{2}{2+k}$.
The spins $\overset{\rightarrow}{S}$ reappear as the boundary conditions
in a boundary CFT (BCFT) represented by the WZW model on $SU(2)$.
Now, the following \emph{fusion rules hypothesis \cite{Affleck95},
}explains the creation and nature of the multichannel Kondo states:
\emph{The infrared fixed point in the $k$-channel spin-$s$ Kondo
problem is given by fusion with the spin-$s$ primary for $s\leqq k/2$
or with the spin $k/2$ primary for }$s>k/2$\emph{.} The level $k$
Kac-Moody algebra, as in the level $k$ WZW $SU(2)$ model, governs
the behavior of the Kondo state in the presence of $k$ channels of
conducting electrons and magnetic impurity of spin $s$. 

Let us now approach this CFT description from the point of view of
4-dimensionsional geometry. Exotic smooth $\mathbb{R}^{4}$'s are
the simplest topological 4-manifolds but without a simple local coordinate
patch presentation. Therefore mathematics in this case should be substantially
different than in other dimensions. In a series of papers the authors
have recently proposed a way to obtain some relative results in this
field \cite{AsselmeyerKrol2009,AsselmeyerKrol2010,Krol2010,Krol2010b,AsselmeyerKrol2011}.
In particular, different (i.e. nondiffeomorphic) so-called small exotic
$\mathbb{R}^{4}$'s correspond to the codimension-1 foliations of
the 3-sphere $S^{3}$ which lies at the boundary of some compact contractable
4-submanifold $A$ in $\mathbb{R}^{4}$. This compactum $A$ is called
\emph{Akbulut cork} and its embedding, together with the attached
\emph{Casson handles}, is responsible for the exotic smoothness of
$\mathbb{R}^{4}$. The continuous family of small exotic $\mathbb{R}^{4}$
is defined by a radius function $\rho:\mathbb{R}^{4}\to[0,+\infty)$
(polar coordinates) so that $\mathbb{R}_{t}^{4}=\rho^{-1}([0,r))$
with $t=1-\frac{1}{r}$. Next, each member of the radial family determines
a codimension-1 foliation of the homology 3-sphere and these are uniquely
determined by a codimension-1 foliation of some ordinary $S^{3}\subset\partial A$
. However, the codimension-1 foliations of the 3-sphere (especially
the foliated cobordism class) are classified by the 3-rd real cohomology
classes $H^{3}(S^{3},\mathbb{R})$. The integer classes in $H^{3}(S^{3},\mathbb{Z})$
are a special case. It uses flat $PSL(2,\mathbb{R})-$bundles over
the space $(S^{2}\setminus\left\{ \mbox{\mbox{k} punctures}\right\} )\times S^{1}$
where the gluing of $k$ solid tori produces a 3-sphere (so-called
Heegard decomposition). Then one obtains the relation\begin{equation}
\frac{1}{(4\pi)^{2}}\langle GV(\mathcal{F}),[S^{3}]\rangle=\frac{1}{(4\pi)^{2}}\,\intop_{S^{3}}GV(\mathcal{F})=\pm(2-k)\label{eq:integer-GV}\end{equation}
in dependence of the orientation of the fundamental class $[S^{3}]$.
Furthermore we can interpret the Godbillon-Vey invariant as WZW term.
For that purpose we use the group structure $SU(2)=S^{3}$ of the
3-sphere $S^{3}$ or better we identify $SU(2)=S^{3}$. Let $g\in SU(2)$
be a unitary matrix with $\det g=-1$. The left invariant 1-form $g^{-1}dg$
generates locally the cotangent space connected to the unit. The forms
$\omega_{k}=Tr((g^{-1}dg)^{k})$ are complex $k-$forms generating
the deRham cohomology of the Lie group. The cohomology classes of
the forms $\omega_{1},\omega_{2}$ vanish and $\omega_{3}\in H^{3}(SU(2),\mathbb{R})$
generates the cohomology group. Then we obtain for the integral\[
\frac{1}{8\pi^{2}}\intop_{S^{3}=SU(2)}\omega_{3}=1\]
of the generator. This integral can be interpreted as winding number
of $g$. Now we consider a smooth map $G:S^{3}\to SU(2)$ with 3-form
$\Omega_{3}=Tr((G^{-1}dG)^{3})$ so that the integral\[
\frac{1}{8\pi^{2}}\intop_{S^{3}=SU(2)}\Omega_{3}=\frac{1}{8\pi^{2}}\intop_{S^{3}}Tr((G^{-1}dG)^{3})\in\mathbb{Z}\]
is the winding number of $G$. Every Godbillon-Vey class with integer
value like (\ref{eq:integer-GV}) is generated by a 3-form $\Omega_{3}$.
Therefore the Godbillon-Vey class is the WZW term of the $SU(2)$.
For integer $H^{3}(S^{3},\mathbb{Z})$ and for the $S^{3}\subset\partial A$
as above, one has the correspondence: 

\begin{equation}
H^{3}(S^{3},\mathbb{Z})\ni h\mapsto\mathbb{R}_{h}^{4}\:.\end{equation}
 Then we arrive at the correspondence \cite{AsselmeyerKrol2009,Krol2010,AssKrol2010ICM}:
\emph{the level $k$ of the WZW model of CFT on $SU(2)$ (determined
by $k[\:]=h\in H^{3}(S^{3},\mathbb{Z})$) corresponds to the exotic
smooth $\mathbb{R}_{k}^{4}$ when $S^{3}=SU(2)$ is the part of the
boundary of the Akbulut cork $A$. }In that way the exotic $\mathbb{R}_{k}^{4}$
determines the fusions in $SU(2)_{k}$ of boundary spin $s$ impurity
with $k$-channel conducting electrons. And conversely, the symmetry
of the $k$-channel Kondo effect results in 4-dimensions in the corresponding
geometry of exotic $\mathbb{R}_{k}^{4}$. 

{\bf Spinor fields from exotic $\mathbb{R}^{4}$}. Formally Einsteins
equation can be written on every smooth 4-manifold. S\l{}adkowski
showed \cite{Sladkowski2001} that Einsteins equation on the empty
(no sources) but exotic smooth $\mathbb{R}^{4}$ allows solutions
which are equivalent to the solutions of the corresponding equation
with matter sources on the standard $\mathbb{R}^{4}$. In this sense
the exoticness of the smooth structure generates matter in spacetime.
The extension of this correspondence to quantum matter would be an
important ingredient of our understanding of the relation of QM and
GR and was considered already from various points of view \cite{Krol:04b,Krol:2005,AsselmeyerRose2010}.

Recent work of one of the authors and Ros\'e for compact 4-manifolds\cite{AsselmeyerRose2010}
shows how to obtain the action terms of fermions and bosons by using
the Einstein-Hilbert action and the Casson handle. The Casson handle
consists of towers of kinky handles, i.e. neighborhoods of disks with
self-intersections. Every kinky handle can be described by an immersion
(injective differential) $D^{2}\times D^{2}\to M$ into a 4-manifold
$M$. Usually the boundary of the immersed disk is a singular knot
$K$ and the image of the immersion is a knotted solid torus $T(K)=K\times D^{2}$
\cite{AsselmeyerRose2010}. Asselmeyer-Maluga and Ros\'e \cite{AsselmeyerRose2010}
gave an alternative method to describe the immersion by using a spinor
$\Psi$. Then it is possible to express the mean curvature $H_{T(K)}$
of the knotted torus $T(K)$ by $2\overline{\Psi}\sigma_{\mu}\Psi$.
But the mean curvature will be used in the boundary term of the Einstein-Hilbert
action. Let $\psi$ be the extension of $\Psi$ to the 4-manifold
then we obtain \cite{AsselmeyerRose2010} the pure fermionic part
of the action:

\begin{equation}
S(M)=\intop_{M}\left(R+\sum_{n}(\psi\cdot\partial_{D}\overline{\psi})_{n}\right)\sqrt{g}d^{4}x\end{equation}
where $R$ is the scalar curvature on $M$ and $\psi$ the spinor
field extended over $M$. Let us note that the calculations relies
merely on the Casson handle structure which means that the fermionic
field fulfilling the Dirac equation appears also in the open 4-manifold
case of exotic $\mathbb{R}^{4}$. In general there are many Casson
handles involved in the description of an exotic $\mathbb{R}^{4}$
which means that several fermionic fields can be assigned to. Besides,
one can decompose a complicated Casson handle into a set of different
Casson handles so the number of the resulting fermionic fields is
a function of this decomposition and the complexity of the Casson
handles. Let us denote these fermionic fields corresponding to the
small exotic $\mathbb{R}_{k}^{4}$ as $\left\{ \psi's\right\} _{k}$.

{\bf The model}. Fix an Akbulut cork $A$ and Casson handle(s) embedded
in $\mathbb{R}^{4}$. Magnetic spin $s$ is associated with the $SU(2)$
symmetry. Let this $SU(2)=S^{3}$ lies at the boundary of $A$. Generate
spinor fields as above: spin-particles (electrons) $\left\{ \psi's\right\} $
from Casson handle(s). They are free, in Minkowski spacetime, as far
as Casson handles are not attached to $A$. When smoothness of $\mathbb{R}^{4}$
is exotic and CH's are attached to $A$ and embedded in $\mathbb{R}^{4}$,
the bound state of electrons and $s$ is thus corresponding to the
exoticness of the smooth structure. The impurity refers to a boundary
state of BCFT and together with channels for spinors they force the
$SU(2)_{k}$ symmetry as before. This in turn determines the class
$k[\:]\in H^{3}(S^{3},\mathbb{Z})$ which corresponds to exotic $\mathbb{R}_{k}^{4}$.
The quantum bound state of electrons and $s$ is the Kondo state.
In case of a single Casson handle a single (non-relativistic) electron
emerges. One can conjecture that the emerging exotic $\mathbb{R}_{1}^{4}$
is the simplest exotic $\mathbb{R}^{4}$ of Bizaca and Gompf in this
case \cite{Asselmeyer2007,GomSti:1999}. The Kondo bound state and
its interactions with conducting electrons are described by boundary
CFT (BCFT) where the impurity is represented by a conformal boundary
conditions of WZW $SU(2)$ at $k=1$.

When we include more than one Cassen handel (CH), say $k$, in the
description of exotic $\mathbb{R}^{4}$ (or a single CH though more
complicated which can be decomposed into a set of CH's) the model
describes the Kondo bound state where the impurity is bounded with
$k$ electrons $\left\{ \psi's\right\} _{k}$ from different conducting
channels. This is again a necessarily quantum process. The value $k$
is a suitable function of the complexity and decomposition of CH.
Thus the geometry of the exotic $\mathbb{R}_{k}^{4}$ describes quantum
entanglement of $k$ conducting electrons from $k$ symmetric channels
with magnetic impurity of spin $s$. When smoothness of $\mathbb{R}^{4}$
is standard, i.e. $k=0\in H^{3}(S^{3},\mathbb{Z})$, the electrons
are free and not bounded by $s$. This means that the proposed model
recognizes the smooth 4-geometry directly from quantum interacting
magnetic matter in spacetime. 

{\bf Discussion}. The presented model assigns the geometry of a 4-dimensional
Euclidean smooth $\mathbb{R}^{4}$ to the quantum interacting spin
matter in the $k$ channel Kondo effect. The basic tool is the WZW
$SU(2)$ model of CFT at the level $k$ which is derived from an exotic
small smooth $\mathbb{R}_{k}^{4}$ and is also crucial in the description
of the $k$ channel Kondo effect. This agreement allows to interpret
fermions generated directly by Casson handles, as in \cite{Asselmeyer-Maluga2010},
as electrons bounded by the magnetic impurity. 

The correlation of 4-geometry and quantum matter occurs at low temperature
and the spin-spin and electromagnetic interactions are much stronger
at the microscopic scale than gravity. The emerging exotic 4-geometry
can be understood as a kind of a relativistic limit for the low energy
quantum matter in SM which incorporates also a gravity component,
i.e. the non-flatness of the exotic $\mathbb{R}^{4}$. Here we do
not uncover the full physical meaning of the correspondence (as e.g.
for cosmology). However this model is the first instance of a possible
modification (in a suitable regime) of a spacetime structure by purely
quantum interactions as in standard model of particles. The question,
whether this modification of a 4-dimensional geometry by quantum spin
matter actually holds in reality and under what physical conditions,
needs further careful studies and model buildings.

Due to the above discussion, the spacetime should be rather modelled
locally by an exotic than a standard $\mathbb{R}^{4}$. The recognition
of the structure of 4-space as replacing quantum interactions, is
an important stage in building the final theory of quantum gravity.

\end{document}